\documentclass{aa}
\usepackage{graphicx}

\newcommand\teff{$ {\rm T_{eff}}$}

\newcommand\logg{$\log {\rm g}$}
\newcommand\loghe{${\rm \log{\frac{n_{He}}{n_{H}}}}$}

\newcommand{\Msolar}{\mbox{\,$\rm M_{\odot}$}}        

\newcommand{\tev}{$\rm{T_{evol}}$}
\newcommand{\tfl}{$\rm{T_{flight}}$}
\newcommand{\hei}{He {\sc i}}

\newcommand{\cii}{C {\sc ii}}
\newcommand{\ciii}{C {\sc iii}}
\newcommand{\nii}{N {\sc ii}}

\newcommand{\oii}{O {\sc ii}}
\newcommand{\nei}{Ne {\sc i}}
\newcommand{\neii}{Ne {\sc ii}}
\newcommand{\mgii}{Mg {\sc ii}}
\newcommand{\alii}{Al {\sc ii}}
\newcommand{\aliii}{Al {\sc iii}}
\newcommand{\silii}{Si {\sc ii}}
\newcommand{\siliii}{Si {\sc iii}}

\newcommand{\piii}{P {\sc iii}}
\newcommand{\sii}{S {\sc ii}}
\newcommand{\siii}{S {\sc iii}}
\newcommand{\arii}{Ar {\sc ii}}

\newcommand{\feiii}{Fe {\sc iii}}

\begin{document}
\title{Early type stars at high galactic latitudes\\
       I: Ten young massive B-type stars
\thanks{Based on observations 
    obtained 
    at the W.M. Keck Observatory, which is operated by the Californian 
    Association for Research in Astronomy for the California Institute of 
    Technology and the University of California
}
\thanks{Based on observations collected at the German-Spanish Astronomical 
Center (DSAZ), Calar Alto, operated by the Max-Planck-Institut f\"ur 
Astronomie Heidelberg jointly with the Spanish National Commission for 
Astronomy} 
\thanks{Based on observations collected at the European Southern
        Observatory (ESO proposal No. 65.H-0341(A))}
}

\author{M. Ramspeck
 \and U. Heber \and S. Moehler}

\offprints{M. Ramspeck}

\institute{Dr.-Remeis-Sternwarte, Universit\"at Erlangen-N\"urnberg,
           Sternwartstr. 7,
           D-96049 Bamberg, Germany
           e-mail: ramspeck@sternwarte.uni-erlangen.de}

\date{received;  accepted }

\abstract{
We present the results of quantitative spectral analyses of ten 
apparently normal B-type stars. These stars were found to be young massive  
B-type stars at distances of z=2.6 to 7.6\,kpc from the galactic plane based on 
their positions 
in the (\teff, \logg) diagram, normal abundance patterns and/or large projected 
rotational velocities. We discuss formation scenarios (runaway star 
scenarios versus a scenario for star formation in the halo) by comparing 
times-of-flight and evolutionary time scales. 
For all stars (except SB~357 and HS~1914$+$7139) both the scales are similar 
indicating that the stars could have 
formed in the galactic disk and been ejected from there soon after 
their birth. 
Derived ejection velocities range from 130\,km s$^{-1}$ to 440\,km s$^{-1}$ and 
may be used to constrain models for ejection mechanisms. 
Using new proper motion measurements we show that PHL~346, 
which was considered the most likely candidate for a young B-type star born in 
the halo, can be explained as a runaway star from the galactic plane.
}
\authorrunning{Ramspeck et al.}
\titlerunning{Early type stars}
\maketitle

\section{Introduction\label{intro}} 
Main sequence B-type stars located far away from the galactic plane are a rare,
albeit known phenomenon. In their pioneering paper, Greenstein \& Sargent
(\cite{grsa74}) studied faint blue stars at high galactic latitudes and
classified 25\% of them as apparently normal OB-type stars at distances from the
galactic plane of z=1--3kpc. More detailed studies have shown that many,
but not all,  of the apparently normal stars were in fact highly evolved
low-mass stars. Surveys for UV excess objects (e.g., Palomar Green,
Hamburg-Schmidt, Edinburgh Cape) have found many new candidates (e.g., 
Saffer et al \cite{sake97}, Rolleston et al. \cite{roha99}, Magee et al.
\cite{madu98}) and some may be known even in other galaxies (M~31, Smoker
et al. \cite{smke00}). Their properties, possible evolutionary histories
and formation mechanisms were reviewed by, e.g.,Tobin (\cite{tobi87}),
Keenan (\cite{keen92}), and Heber et al. (\cite{hemo97}). 

Tobin (\cite{tobi87}) also discusses the problem that some highly evolved
stars spectroscopically mimic massive stars almost perfectly. The
most striking example is PG~0832+676 which has been analysed several
times. Its abundance pattern is close to normal. Only recently, Hambly et
al. (1996) were able to firmly establish slight underabundances and a very
low projected rotation velocity. Combining both results they
concluded that PG~0832+676 in fact is a highly evolved
star. Abundance analyses as well as determinations of rotational velocities
are thus of essential importance for the verification of massive B-type star
candidates. A high rotational velocity generally excludes a late
evolutionary status of the star, as old, low-mass stars cannot rotate as fast
as massive stars.
This fact was used, e.g., by Heber et al. (\cite{hemo95}, HS~1914$+$7139)
and Schmidt et al. (\cite{scde96}, PG~0009$+$036) to identify massive 
B-type stars far from the Galactic plane from medium-resolution spectra. 

The massive B-type stars in galactic halos can be separated kinematically into two
different categories: those stars with a main sequence
lifetime larger than the time they would need to travel from the plane to
their present position and those with a main sequence lifetime too small to
reach their current position assuming an acceptable velocity vertical to the
galactic disk. The former ones are assumed to be born in the disk and
thereafter ejected from it (runaway stars), while the latter are supposed to be born in the
halo (see Conlon et al. \cite{cobr88}, \cite{codu90}, Hambly et al.
\cite{haco93} for more details). Since many years runaway stars are known 
to exist, whereas formation of massive stars in the halo has not yet 
been confirmed convincingly. The best studied candidate is PHL~346 (Ryans 
et al. \cite{ry96}), a $\beta$ Cephei star in the halo (Dufton et al. 
\cite{du98}). A search for coeval stars around  
PHL~346 (Hambly et al. \cite{hamb96}) met with limited success, since only one 
out of 16 A- and B-type stars around 
PHL~346 was found to have the appropriate spectral type and radial velocity.

Calculations of galactic orbits are thus
very important to determine the true nature of the stars, since they also
allow to determine ejection velocities from the galactic disk. However,
accurate proper motions are a prerequisite for such an analysis. A big
step forward has been achieved by Thejll et al. (\cite{thfl97}) and the 
Hipparcos/Tycho mission (Perryman et al. \cite{peli97}; H{\o}g et al. 
\cite{hofa00}).

In this paper we present the analysis of new high-resolution spectra for 10
apparently normal B-type stars. Half of the sample are new discoveries whereas 
the other half has already been studied previously
(Conlon et al. \cite{codu92}, Ryans et al. \cite{ry96}, Rolleston et al. 
\cite{roha99}, Heber et al. \cite{hemo95}). For the latter the times-of-flight 
quoted in the literature appear to be larger than the evolutionary 
times indicating they might have formed in the halo. 
However, proper motions were not available rendering these estimates 
of the times-of-flight uncertain. Since proper motion measurements 
became available recently for three of these stars, it was deemed necessary to 
reanalyse them from new high resolution spectra.

\section{Observations and Data Reduction}
We have obtained high resolution spectra for all programme 
stars using the HIRES spectrograph at the Keck I telescope, the FEROS 
spectrograph at the ESO 1.5m telescope, the FOCES spectrograph at the 
DSAZ 2.2m telescope and the CASPEC spectrograph at the ESO 3.6m telescope 
(see Table~\ref{hres}). Since the normalization procedure for the Echelle 
spectra is cumbersome for the rather broad Balmer lines 
(see below) it was deemed necessary to secure low resolution spectra at 
least for some of the programme
stars to obtain independent estimates of the atmospheric parameters, in 
particular of the gravities. Appropriate spectra were obtained at Calar 
Alto and at ESO.   
Details are given in Table \ref{hres}.
For PG~1511$+$367, PG~1533$+$467 and PG~1610$+$239 we have new low resolution 
spectra presented in this paper. These spectra were taken with the TWIN spectrograph 
at the DSAZ 3.5m telescope and reduced as described in Edelmann et al. (\cite{edhe01}).
The low resolution spectrum of PG~0122$+$214 were
reduced as described in Moehler et al. (\cite{mohe97}).

For the Keck-HIRES spectra observed in 1998 the blue cross disperser was used
and the spectra cover the blue spectral range (3600 to 5130 \AA).
The standard data reduction as described by 
Zuckerman \& Reid (\cite{Zuc98}) 
resulted in spectral orders that have a somewhat wavy continuum. 
Due to the merging of the higher Balmer lines the removal of the remaining 
waviness is difficult.
We used the spectrum of H1504$+$65 
(a very hot pre-white dwarf devoid of hydrogen and helium, Werner 
\cite{Wer91}),
which was observed in the same night, for 
rectification of our spectra. Its spectrum has only few weak lines 
of highly ionized metals in the blue (3600--4480\AA) where the strong Balmer 
lines are found in the B-type stars. Therefore we normalized 
individual spectral orders 1 to 20 (3600--4480\AA) of the B-type stars by
dividing through the smoothed spectrum of H1504$+$65. The remaining 
orders were normalized by fitting the continuum with spline functions
(interpolated for orders 26 and 27 which contain H$\beta$). 
Judging from the match of line profiles in 
the overlapping parts of neighboring orders this procedure worked 
extremely well. 
In 1996 the red cross disperser of the Keck-HIRES spectrograph was used and 
therefore the blue
part of the spectrum shortward of 4200\AA\ was not recorded. 
Since no merging Balmer lines are present in this spectral range the waviness 
of the spectrum could be 
removed by fitting the continuum with spline functions. For orders 
containing broad Balmer lines the fit functions were interpolated between
neighbouring orders.
   
The FOCES spectra were reduced as described in Pfeiffer et al. 
(\cite{pffr98}) with a software package 
developed by the Munich Group. For the FEROS data
the MIDAS reduction pipeline (Fran\c{c}ois, \cite{fr99}) was used.
For CASPEC data the procedure described by Heber et al. (\cite{hebe86}) was 
applied. The CASPEC and FEROS spectra were normalized in a similar way as 
described for the Keck HIRES spectra of 1996.\\ 
Due to an error in the wavelength calibration no reliable radial velocity 
could be measured for HS~1914$+$7139.
\begin{table*}
\centering
\caption[]{Observational parameters}
\label{hres}
\begin{tabular}{|c|cccc|} \hline
 Name & Observation Date \& & Telescope \& & Resolution & 
 Wavelength Range  \\ 
      & Time (UT)  &  Instrument & (FWHM \AA) &  (\AA)  \\ \hline
      & \multicolumn{4}{c|}{Echelle spectra:} \\
 PG~0122$+$214    & Jul 20, 1998 14:48 & Keck HIRES            & 0.09 & 3600 -- 5130  \\
 PG~1511$+$367    & Jul 20, 1999 20:30 & Calar Alto 2.2m FOCES & 0.15 & 3890 -- 6995  \\
 PG~1533$+$467    & Jul 19, 1999 21:00 & Calar Alto 2.2m FOCES & 0.15 & 3890 -- 6995  \\
 PG~1610$+$239    & Jul 24, 1996   --  & Keck HIRES            & 0.09 & 4265 -- 6720  \\
 PG~2219$+$094    & Jul 20, 1999 03:00 & Calar Alto 2.2m FOCES & 0.15 & 3890 -- 6995  \\
 PHL~159          & Sep 12, 1998 22:00 & Calar Alto 2.2m FOCES & 0.15 & 3870 -- 6830  \\
 PHL~346          & Oct 19, 1986 22:50 & ESO 3.6m CASPEC       & 0.20 & 4070 -- 5130 \\
 SB~357           & Oct 1984           & ESO 3.6m CASPEC       & 0.20 & 4060 -- 5090 \\
 HS~1914+7139     & Jun 05, 1996 13:08 & Keck HIRES            & 0.09 & 4280 -- 6720  \\
 BD$-$15$^\circ$115 & Sep 08, 2000 07:49 & ESO 1.5m FEROS       & 0.09 & 3630 -- 8860  \\ \hline
      & \multicolumn{4}{c|}{low resolution spectra:} \\
 PG~0122$+$214 & Aug 07, 1990 02:32 & Calar Alto TWIN 3.5m & 3.5 & 3875 -- 5010  \\
 PG~1511$+$367 & Jul 18, 1999 20:49 & Calar Alto TWIN 3.5m & 2.9 & 3350 -- 7700  \\
 PG~1533$+$467 & Jul 18, 1999 21:00 & Calar Alto TWIN 3.5m & 2.9 & 3350 -- 7700  \\
 PG~1610$+$239 & Apr 11, 2001       & Calar Alto TWIN 3.5m & 1.0 & 4100 -- 4950  \\
 PG~2219$+$094$^\star$ & Jun 10, 1987 08:35 & ESO MPIA 2.2m B\&C   & 2.5 & 4035 -- 4900  \\ \hline
\end{tabular}\\[2mm]
\begin{tabular}{ll}
\parbox[t]{150mm}{ $^\star$ Moehler et al. (1990)}
\end{tabular}\\
\end{table*}

\begin{figure*}
\vspace{12.0cm}
\includegraphics{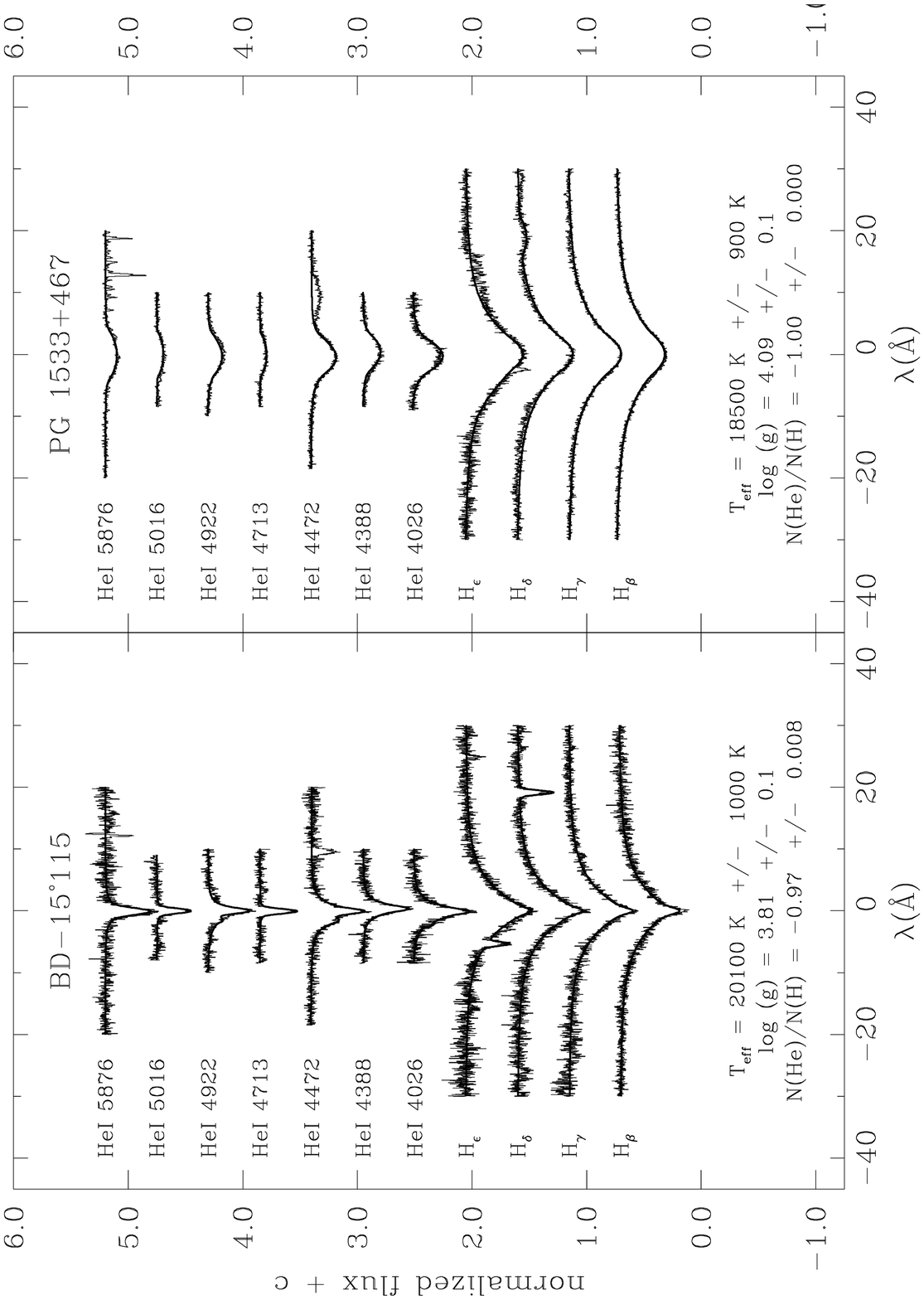}
	\caption[]{Fit examples for a slowly rotating star (BD$-$15$^\circ$115, left hand side)
                   and a rapidly rotating star (PG~1533$+$467, right hand side) 
 	\label{fit}
}
\end{figure*}
\section{Atmospheric Parameters and Projected Rotational Velocities}
To derive atmospheric parameters (effective temperature, surface gravity,
and photospheric helium abundance) and projected rotational velocities all 
Balmer lines
and the He~{\sc i} lines $\lambda\lambda$ 4026~\AA, 4388 \AA, 4438 \AA, 
4472 \AA, 4713 \AA, 4922~\AA, 5016\AA, 5048 \AA, 5678 \AA, 
in the observed spectra were fitted with synthetic line 
profiles calculated from model atmospheres. 

We computed LTE model atmospheres using
the program of Heber et al. ({\cite{hebe00}), which calculates 
plane parallel, chemically 
homogeneous and fully line blanketed models, using the 
opacity distribution functions for metal 
line blanketing by Kurucz (\cite{kuru79}, ATLAS6). 
From these model atmospheres synthetic spectra were calculated with
Lemke's version\footnote{For a description see
http://a400.sternwarte.uni-erlangen.de/$\sim$ai26/linfit/linfor.html} of
the LINFOR program (developed originally by Holweger, Steffen, and
Steenbock at Kiel University). The spectra 
include the Balmer lines H$_\alpha$ to H$_{22}$ and the \ion{He}{i} lines listed 
above
and the grid covers the range 11\,000 K$\leq$~\teff~$\leq$ 40\,000 K,
3.5 $\leq$~\logg~$\leq$ 6.5 and --4.0 $\leq$~\loghe~$\leq$ --0.5 at solar 
metallicity.

The fit procedure is based  on a $\chi^{2}$ test using the routines 
developed by Bergeron et al. (\cite{besa92}) and Saffer et al.
(\cite{sabe94}) and modified by Heber et al. (\cite{hena97}) 
to derive also the rotational
velocity.  
The theoretical spectra are convolved with the
instrument profiles (Gaussian with the appropriate instrumental FWHM) and a 
rotational profile. The fit program then normalizes theoretical and 
observed spectra using the same continuum points. 
Example fits for a rapidly rotating and a slowly rotating programme star are 
shown in Fig.\ref{fit} for hydrogen and helium lines, whilst 
Fig.\ref{compstars} compares the metal line spectra 
of slowly rotating (PHL~159, BD$-$15$^\circ$115) stars
and a rapidly rotating star (PG~1533$+$467).
\begin{figure*}
\vspace{12.0cm}
\includegraphics{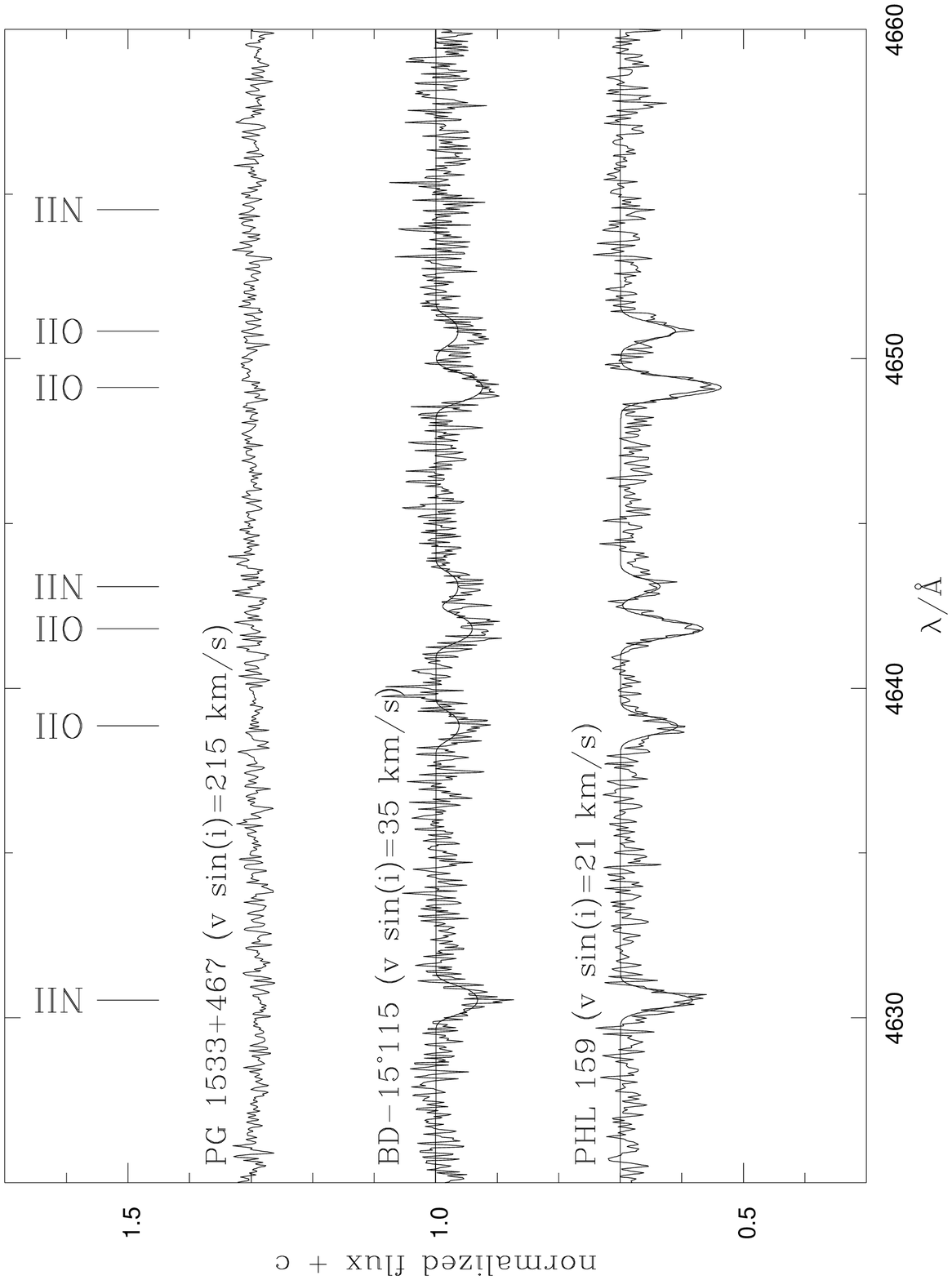}
	\caption[]{Wavelength range with strong {\nii} and {\oii} lines to show examples 
                   for spectra of slowly rotating (bottom, middle) and rapidly rotating
                   (top) stars.
 	\label{compstars}
}
\end{figure*}
\begin{figure}
\vspace{8.5cm}
\includegraphics{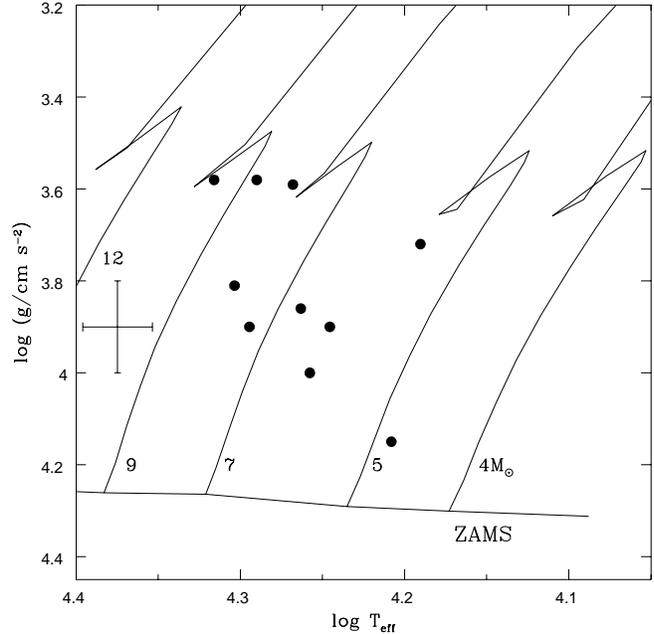}
 \caption[]{Positions of the programme stars (filled circles) in a (\teff, \logg) diagram
with evolutionary tracks calculated by Schaller et al.(1992) for 
determining the masses and evolution times.
\label{schaller}
}
\end{figure}
\begin{table*}
\centering
\caption[]{Atmospheric parameters and rotational velocities for the progamme
stars as derived from high and low resolution spectroscopic data and 
comparison of these data with the effective temperatures calculated from
Str{\o}mgren photometry. 
The rotational velocities derived from the high
resolution data were used to fit the low resolution spectra.\\
$^\star$: \teff, \logg{ }fixed and He fitted. $^\star$$^\star$: 
Helium is fixed at solar abundance, see text.\\
For the low resolution data He is fixed at solar abundance.}
\label{comparison}
\begin{tabular}{|c|cccc|cc|cc|} \hline
      &   \multicolumn{4}{c|}{High Resolution} & \multicolumn{2}{c|}{Low Resolution}  & \multicolumn{2}{c|}{Photometry} \\
 Name & $\rm{T_{eff}}$  & $\rm{\log(\frac{g}{cm s^{-2}})}$ & 
 $\rm{\log\frac{N(He)}{N(H)}}$ & $v \sin i$  &   
 $\rm{T_{eff}}$ & $\rm{\log(\frac{g}{cm s^{-2}})}$ & 
 $\rm{T_{eff}}$ & E(b--y) \\ 
 & (K) &  &  &  (km s$^{-1}$) & (K) &  & (K) & \\ \hline
 PG~0122$+$214 & 18\,300 & 3.86 & --0.98 & 117 & 18\,700 & 3.90 &  18\,500 (1) & 0.0 \\
 PG~1511$+$367 & 16\,100 & 4.15 & --1.16 & 77 & 15\,600 & 4.20  &  15\,900 (1) & 0.0 \\
 PG~1533$+$467 & 18\,500 & 4.09 & --0.94$^\star$ & 215 & 17\,700 & 3.93 &  17\,700 (1) & 0.020 \\
 PG~1610$+$239 & 15\,500 & 3.72 & --0.84$^\star$ &  75 & 15\,400 & 3.69 &  18\,600 (1) & 0.082 \\ 
 PG~2219$+$094 & 19\,500 & 3.58 & --1.00$^\star$ & 225 & 18\,200 & 3.52 &  16\,700 (2) & 0.037 \\
               &         &      &                &     &         &      & 19\,500 (3)    & 0.081 \\
 PHL~159 & 18\,500 & 3.59 & --0.84 & 21 & -- & -- & 20\,900 (4) & 0.025 \\
 PHL~346         & 20\,700 & 3.58 & --1.00 & 45 & -- & -- &  22\,300  (7)   & 0.037 \\ 
 SB~357          & 19\,700 & 3.90 & --1.00$^\star$$^\star$ & 180 & -- & -- & 19\,700 (5)   & 0.052 \\ 
                 &         &      &                        &     &    &    & 19\,700 (8)   & 0.061 \\
                 &         &      &                        &     &    &    & 19\,800 (9)   & 0.037 \\ 
 BD$-$15$^\circ$115 & 20\,100 & 3.81 & --0.97   & 35 & -- & -- &   19\,800 (5)  & 0.0 \\ 
                    &         &      &          &    &    &    &   20\,200 (6) & 0.0 \\ 
 HS~1914$+$7139 & 17\,600 & 3.90 & --0.99 & 250 & 18\,100 & 3.60 & -- & -- \\ \hline
\end{tabular} \label{fitres}\\[2mm]
\parbox[t]{150mm}{\it{References}: 
(1) Wesemael et al. (\cite{wefo92}); (2) Mooney et al. (\cite{moro00}); 
(3) Moehler et al. (\cite{mori90});\\  
(4) Brown et al (\cite{brki79});
(5) Hauck \& Mermilliod (\cite{hame98}); (6) Kilkenny et al. (\cite{kihi75});\\
(7) Kilkenny et al. (\cite{kihi77}); (8) Kilkenny D. (\cite{kil95});
(9) Graham et al. (\cite{grsl73})}
\end{table*}
The fit procedure was executed for all high and low resolution spectra and the
results are listed in Table \ref{comparison}\footnote{For the fit of the low 
resolution spectra the He abundance was kept fixed at the solar value and the 
projected rotation velocity derived from the high resolution spectra were used.}. 
Formal fitting errors are very small for the high resolution spectra, 
on average $\Delta \rm{T_{eff}=100 K}$, $\Delta \rm{\log g=0.02}$.
Systematic errors (e.g. continuum placement, uncertainties in line broadening
theory)
are certainly larger and therefore dominate the error budget. 
We estimate errors in effective temperatures
conservatively as 5\% and adopted an error of $\pm$0.1\,dex for the
gravities. 
The fitting errors for the Helium abundance are on average 
$\Delta \rm{\log (n(He)/n(H))=0.05}$. Since sharp Helium lines as well 
as broad Helium lines are well reproduced (see Fig. \ref{fit}) systematic
errors due to {\hei} line broadening theory appear to be small and
we adopted an error of $\pm$0.1\,dex in all cases.

For rapidly rotating stars the $\chi^2$ minimum 
is too poorly defined to allow a reliable determination of the He abundance
simultaneously. Therefore, in a first step the 
helium abundance was kept fixed at $-$1.00 (i.e. solar) for the fit procedure.
In a second iteration step the helium abundance was determined by fitting  
the helium lines while keeping the \teff{ }and 
\logg{ }fixed at those values determined in the first iteration step.
For all stars (except HS 1914$+$7139) Str{\o}mgren photometry 
is available, which allowed an independent determination of the effective 
temperature. We used the program of
Moon (\cite{moon85}) as modified by Napiwotzki et al. (\cite{nasc93}) to derive
the effective temperature and the reddening and 
compare the photometric temperatures to the spectroscopic ones in 
Table~\ref{comparison}. There is a good agreement between results from low
and high resolution spectra and photometry, except for PG~1610$+$239, 
PHL~159 and PHL~346.
The spectrum of SB~357 shows the presence of emission in $\rm{H_{\beta}}$ and
$\rm{H_{\gamma}}$ but not in $\rm{H_{\delta}}$. Therefore the effective temperature
were obtained from Str{\o}mgren photometry and the surface
gravity from fitting the far wings of the hydrogen lines. The helium lines of 
this object were difficult to fit, but the observation is compatible with normal
abundance and there is no indication of emission in any of the helium lines observed.
The parameters used for further analyses were taken from the high resolution spectra,
because of the larger wavelength coverage and the excellent quality of the fits.
In the case of PG~1533$+$467, however, the wavelength coverage of the low resolution
spectrum is larger than that of the high resolution one and 
therefore we used the average. The finally adopted parameters are listed 
in Table~\ref{results}.
Results are shown in a (\teff, \logg) diagram (Fig. \ref{schaller}).

\section{Chemical Abundances}

Seven programme stars (PG~0122$+$214, PG~1511$+$367, PG~1533$+$467, 
PG~1610$+$239, PG~2219$+$094, HS~1914$+$7139 and SB~357) display highly 
broadened lines 
(due to rotation, see Table \ref{comparison}). Only the strongest metal 
lines (e.g. {\cii} 4267\AA, {\mgii} 4481\AA) could be identified. Therefore 
it was impossible to perform a detailed abundance analysis.

The equivalent widths were measured employing the nonlinear least-squares
Gaussian fitting routines in MIDAS with central wavelength, central 
intensity and full width at half maximum as adjustable parameters.
For metal lines located in the wings of
Balmer or helium lines an additional Lorentzian function is 
used to describe the line wings of the latter.

Metal lines of the species {\cii}, {\ciii}, {\nii}, {\oii},
Ne~{\sc i}, Mg~{\sc ii}, {\alii}, {\aliii}, {\silii}, {\siliii},
{\piii}, {\sii}, {\siii}, {\arii} and {\feiii} 
were identified in the sharp-lined spectra of BD$-$15$^\circ$115, PHL~159 and 
PHL~346. The atomic data for the analysis were taken from several 
tables: 
\begin{enumerate}
 \item CNO from Wiese et al. (\cite{wifu96})
 \item Fe from Kurucz (\cite{kuru92}) and Ekberg (\cite{ekbe93})
 \item Ne, Mg, Al, Si, S, P, Ar from Hirata et al. (\cite{hiho95})
\end{enumerate}
\begin{figure*}
\vspace{8.0cm}
\includegraphics{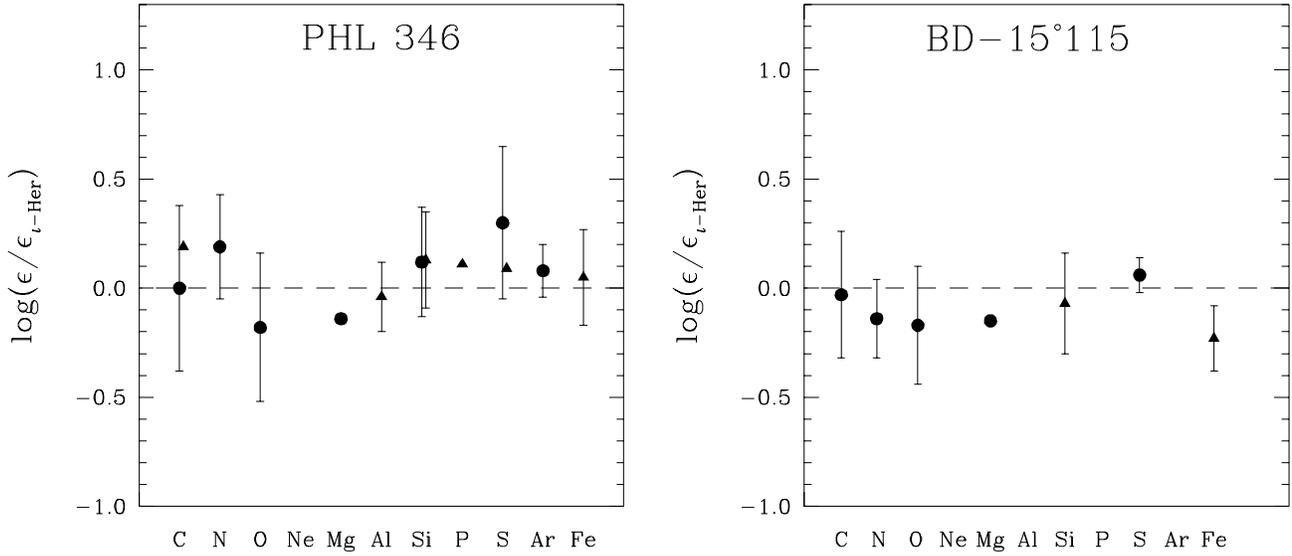}
 \caption[]{LTE abundances (relative to $\iota$~Her) and errors 
of the programme stars. Abundances derived from singly ionized elements are shown 
as filled circles and from doubly ionized ones as filled triangles. 
 \label{metal}
}
\end{figure*}
\begin{figure}
\vspace{8.3cm}
\includegraphics{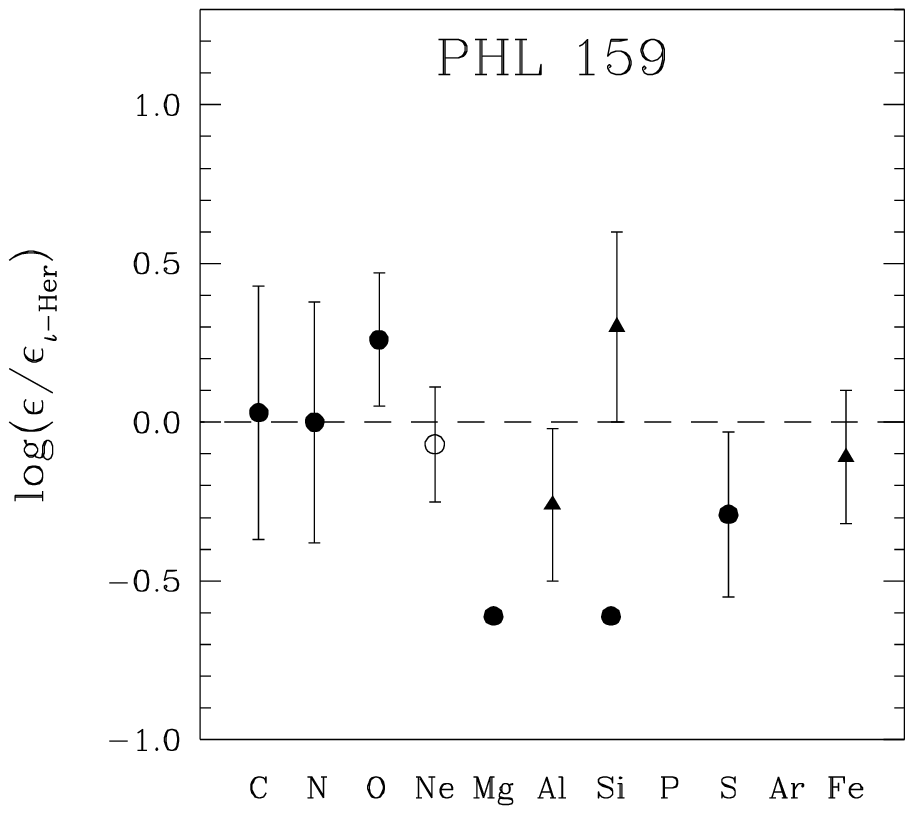}
 \caption[]{Like Fig. \ref{metal}: LTE abundances (relative to $\iota$~Her) 
for PHL~159.\\ Abundances derived from neutral elements are shown as open circles, 
from singly ionized ones as filled circles and from doubly ionized ones as filled triangles. 
 \label{metal2}
}
\end{figure}

The LTE abundances were derived by using the classical curve--of--growth
method and the LINFOR program (version of Lemke, see above). 
In this case the model
atmospheres were generated for the appropriate values of effective
temperature, gravity and solar helium and metal abundance with the 
ATLAS9 program of Kurucz (\cite{kuru92}).\\ 

Then we calculated curves of growth for the observed metal lines, from which
abundances were derived. 
Blends from different ions were omitted from the analysis. In the final 
step the abundances were determined from a detailed spectrum synthesis 
(using the LINFOR code described above) of 
all lines measured before. 
The results of the LTE abundance analysis and the r.m.s. errors for
PHL~346 and BD$-$15$^\circ$115 are shown in Table \ref{LTEcomp}  
and compared with other analyses and for PHL~159 in Table \ref{LTEphl159}. 
Besides the statistical r.m.s. errors (given in Tables \ref{LTEphl159} and 
\ref{LTEcomp}) the uncertainties in \teff, \logg\ and microturbulent 
velocity (see below) contribute to the error budget.
In order to minimize the systematic errors we use the B-type star $\iota$~Her 
as a comparison star. This star has been analysed by Hambly et al. (\cite{haro97}).
We redetermined the LTE abundances of $\iota$~Her using 
the same atomic data, model atmosphere and spectrum synthesis code as for our 
programme stars and took the equivalent widths measured by Hambly et al. (\cite{haro97}).\\
Our results for $\iota$~Her agree to within 0.1\,dex with those of Hambly et al.
(\cite{haro97}) except for {\cii} (0.12\,dex), {\siliii} (0.17\,dex), {\siii} (0.21\,dex)
and {\feiii} (0.36\,dex). In particular our
statistical error for {\feiii} is much lower than that of Hambly et al. (\cite{haro97}).
These differences can be attributed to different oscillator strengths used.\\ 
Results are given in Tables \ref{LTEphl159} and \ref{LTEcomp} and systematic errors
are adopted for our programme stars as well. These errors are incorporated 
in the error bars plotted in Figs. \ref{metal} and \ref{metal2}.
\begin{table*}
\centering
\caption[]{LTE abundances of PHL~159 compared with $\iota$~Her as a 
            comparison star and the range of LTE abundances from 21 B-type stars
            analysed by Kilian (\cite{kili94}). 
            For $\iota$~Her the abundances are determined with the equivalent widths
            from Hambly et al. (\cite{haro97}). For $\iota$~Her systematic errors due 
            to uncertainties of atmospheric parameters have been determined in this work and 
            are listed in parentheses.}
\label{LTEphl159}
\begin{tabular}{|l|r@{$\,\pm\,$}rr|r@{$\,\pm\,$}rr|r@{$\,\pm\,$}rr} \hline
     Element
      & \multicolumn{3}{c|}{$\iota$~Her}
      & \multicolumn{3}{c|}{PHL~159} 
      & \multicolumn{3}{c|}{B-type stars} \\
      & \multicolumn{3}{c|}{} & \multicolumn{3}{c|}{this paper} & \multicolumn{3}{c|}{K94} \\ \hline
$\xi$ (km s$^{-1}$)  & \multicolumn{3}{c|}{5} & \multicolumn{3}{c|}{8}      & \multicolumn{3}{c|}{} \\ 
He {\sc i}    & \multicolumn{3}{c|}{10.78} & 11.16 & 0.10 & (5)             & \multicolumn{3}{c|}{} \\
C  {\sc ii}   & 8.14 & 0.26 & ($\pm$0.21)  & 8.17 & 0.35 & (5)              & \multicolumn{3}{c|}{8.02--8.95} \\
N  {\sc ii}   & 7.85 & 0.18 & ($\pm$0.14)  & 7.85 & 0.36 & (9)              & \multicolumn{3}{c|}{7.48--8.30} \\
O {\sc ii}    & 8.72 & 0.16 & ($\pm$0.19)  & 8.98 & 0.17 & (17)             & \multicolumn{3}{c|}{8.24--8.65} \\
Ne {\sc i}    & \multicolumn{3}{c|}{}      & 8.49 & 0.18 & (13)             & \multicolumn{3}{c|}{} \\
Ne {\sc ii}   & \multicolumn{3}{c|}{}      & \multicolumn{3}{c|}{--}        & \multicolumn{3}{c|}{7.99--8.19$^{(1)}$} \\
Mg {\sc ii}   & 7.28 & 0.20 & ($\pm$0.12)  & \multicolumn{2}{c}{6.67} & (1) & \multicolumn{3}{c|}{7.02--7.68} \\
Al {\sc ii}   & \multicolumn{3}{c|}{6.18}  & \multicolumn{3}{c|}{--}        & \multicolumn{3}{c|}{} \\
Al {\sc iii } & 6.31 & 0.12 & ($\pm$0.14)  & 6.05 & 0.20 & (2)              & \multicolumn{3}{c|}{5.71--6.36} \\
Si {\sc ii}   & 6.86 & 0.50 & ($\pm$0.18)  & \multicolumn{2}{c}{6.38} & (1) & \multicolumn{3}{c|}{} \\
Si {\sc iii } & 7.34 & 0.20 & ($\pm$0.21)  & 7.64 & 0.24 & (4)              & \multicolumn{3}{c|}{6.73--7.65} \\
P  {\sc iii } & \multicolumn{3}{c|}{5.53}  & \multicolumn{3}{c|}{--}        & \multicolumn{3}{c|}{} \\
S {\sc ii}    & 6.99 & 0.19 & ($\pm$0.05)  & 6.70 & 0.26 & (4)              & \multicolumn{3}{c|}{} \\
S {\sc iii }  & 6.93 & 0.32 & ($\pm$0.17)  & \multicolumn{3}{c|}{--}        & \multicolumn{3}{c|}{6.23--7.48} \\
Ar {\sc ii}   & 6.64 & 0.30 & ($\pm$0.08)  & \multicolumn{3}{c|}{--}        & \multicolumn{3}{c|}{} \\
Fe {\sc iii } & 7.33 & 0.09 & ($\pm$0.15)  & 7.22 & 0.15 & (2)              & \multicolumn{3}{c|}{7.09--7.80} \\ \hline
\end{tabular}\\[2mm]
\begin{tabular}{ll}
\parbox[t]{160mm}{\it{References}: K94=Kilian (\cite{kili94}) \\
                  (1) $=$ see text for a discussion of the Ne abundances}
\end{tabular}
\end{table*}           
\begin{table*}
\centering
\caption[]{Comparison of LTE abundances for 
           PHL~346 and BD$-$15$^\circ$115 
           with results from literature. The number of spectral lines used is given in brackets.
           Errors for the programme stars are statistical errors only.}
           
\label{LTEcomp}
\begin{tabular}{|l|r@{$\,\pm\,$}rr||r@{$\,\pm\,$}rr|r@{$\,\pm\,$}rr||r@{$\,\pm\,$}rr|r@{$\,\pm\,$}rr|} \hline
     Element
      & \multicolumn{3}{c||}{B-type stars}
      & \multicolumn{3}{c|}{PHL~346}
      & \multicolumn{3}{c||}{PHL~346}
      & \multicolumn{3}{c|}{BD$-$15$^\circ$115} 
      & \multicolumn{3}{c|}{BD$-$15$^\circ$115} \\ 
      & \multicolumn{3}{c||}{K94} &   \multicolumn{3}{c|}{this paper} & 
\multicolumn{3}{c||}{R96} & \multicolumn{3}{c|}{this paper} & 
\multicolumn{3}{c|}{C92} \\ \hline
 $\xi$ (km s$^{-1}$)  & \multicolumn{3}{c||}{} &  \multicolumn{3}{c|}{23} & \multicolumn{3}{c||}{15}   & \multicolumn{3}{c|}{8}   & \multicolumn{3}{c|}{5} \\
 He {\sc i}    & \multicolumn{3}{c||}{} &   11.00 & 0.10  & (6) & 11.05 & 0.09 & (4)        & 11.03 & 0.10 & (8)        & 10.90 & 0.10 & (8) \\ 
 C  {\sc ii}   & \multicolumn{3}{c||}{8.02--8.95}  &  8.14 & 0.32 & (2)      &  8.16 & 0.29 & (3)       &  8.11 & 0.20 & (4)        &  8.00 & 0.30 & (3) \\
 C  {\sc iii}  & \multicolumn{3}{c||}{}     &  \multicolumn{2}{c}{8.33} & (1) & \multicolumn{3}{c||}{--} & \multicolumn{3}{c|}{--} & \multicolumn{3}{c|}{--} \\
 N  {\sc ii}   & \multicolumn{3}{c||}{7.48--8.30}  &  8.04 & 0.19 & (24)     &  8.04 & 0.24 & (19)       &  7.71 & 0.11 & (5)        &  \multicolumn{2}{c}{7.80} & (1) \\
 O {\sc ii}    & \multicolumn{3}{c||}{8.24--8.65}  &  8.54 & 0.28 & (22)     &  8.79 & 0.25 & (42)       &  8.55 & 0.20 & (8)        &  8.80 & 0.30 & (9) \\
 Mg {\sc ii}   & \multicolumn{3}{c||}{7.02--7.68}  &  \multicolumn{2}{c}{7.14} & (1) & \multicolumn{2}{c}{7.37} & (1) & \multicolumn{2}{c}{7.13} & (1) & \multicolumn{2}{c}{7.10} & (1) \\
 Al {\sc ii}   & \multicolumn{3}{c||}{} &  \multicolumn{3}{c|}{--}     & \multicolumn{3}{c||}{--}        & \multicolumn{3}{c|}{--}    &  6.30 & 0.10 &(2) \\
 Al{\sc iii }  & \multicolumn{3}{c||}{5.71--6.36}  &  6.27 & 0.07 & (3)       & 6.20 & 0.06 & (3)        & \multicolumn{3}{c|}{--}    & \multicolumn{3}{c|}{--} \\
 Si {\sc ii}   & \multicolumn{3}{c||}{}  &  6.98 & 0.17 & (3)       &  7.46 & 0.01 & (2)        & \multicolumn{3}{c|}{--}    & \multicolumn{3}{c|}{--} \\
 Si {\sc iii } & \multicolumn{3}{c||}{6.73--7.65}  &  7.47 & 0.08 & (4)       &  7.60 & 0.04 & (2)        &  7.27 & 0.11 & (3)        &  7.50 & 0.30 & (2) \\
 P {\sc iii}   & \multicolumn{3}{c||}{} &  \multicolumn{2}{c}{5.64}  & (1)     &  \multicolumn{2}{c}{5.39}  & (1) & \multicolumn{3}{c|}{--}  & \multicolumn{3}{c|}{--}  \\
 S {\sc ii}    & \multicolumn{3}{c||}{} &  7.29 & 0.35 & (6)       &  \multicolumn{3}{c||}{--}   &  7.05 & 0.06 & (2) &  6.80 & 0.10 & (2) \\
 S {\sc iii }  & \multicolumn{3}{c||}{6.23--7.48} &  \multicolumn{2}{c}{7.02}  & (1) & 7.20 & 0.09 & (7) & \multicolumn{3}{c|}{--}  & \multicolumn{3}{c|}{--} \\
 Ar {\sc ii}   & \multicolumn{3}{c||}{}  &  6.72 & 0.09 & (2)       & \multicolumn{3}{c||}{--}   & \multicolumn{3}{c|}{--}    & \multicolumn{3}{c|}{--} \\
 Fe {\sc iii } & \multicolumn{3}{c||}{7.09--7.80} &  7.38 & 0.16 & (6)       & 6.66 & 0.75 & (7) &  7.10 & 0.04 & (2) &  7.20 & 0.40 & (3) \\ \hline
\end{tabular}\\[2mm]
\begin{tabular}{ll}
\parbox[t]{160mm}{\it{References}: K94=Kilian (\cite{kili94}); 
R96=Ryans et al. (\cite{ry96}); 
C92=Conlon et al. (\cite{codu92})}
\end{tabular}
\end{table*}
The determination of elemental abundances is interlocked with the microturbulent
velocity $\xi$. This can be derived if a sufficient number of lines of one ion
can be measured over a wide range of line strengths. In our programme stars
{\nii} and {\oii} lines are most suitable for this purpose since 
many lines of these ions can be identified. 
Microturbulent velocities of $\xi$ = 8 km $\rm{s^{-1}}$ were found for PHL~159
and BD$-$15$^\circ$115, while a rather high value of $\xi$~= 23~km $\rm{s^{-1}}$ was 
deduced for PHL~346. Our results for PHL~346 and BD$-$15$^\circ$115 are somewhat 
larger than those derived by Ryans et al. (\cite{ry96}) and Conlon et al. 
(\cite{codu92}), see Table \ref{LTEcomp}.

Remarkable is the large difference ($\approx$ 1.0\,dex) between the {\silii} 
and {\siliii} abundances. This has been found in several 
analyses of the comparison star $\iota$~Her as well 
(Hambly~et~al.~, \cite{haro97}; 0.67\,dex). 
In a differential analysis these systematic errors cancel to a large extent.
NLTE effects are small for all elements ($\le$0.1\,dex, Kilian \cite{kili94}) 
except for {\nei}. As demonstrated by Auer \& Mihalas (\cite{aumi73}) LTE 
calculations overestimate the neon abundance. They
carried out NLTE 
calculations for {\nei} in $\iota$~Her and derived a neon abundance (close 
to solar) which is lower by 
0.60~dex than our LTE result. Therefore our absolute Ne abundances are 
overestimated. 
The abundances of the programme stars with respect to 
$\iota$~Her are plotted in Fig.~\ref{metal}.

\begin{enumerate}
 \item {\it{BD$-$15$^\circ$115:}}\\
        All abundances are in good agreement with 
        $\iota$~Her (to within error limits) except for Mg and Fe 
        which are underabundant by about 0.2\,dex. Our results are in accordance 
        with those of a previous analysis of the star by Conlon et al.
        (1992, see Table~\ref{LTEcomp}), except for Si for which the 
        authors derive a considerably larger abundance.
 \item {\it{PHL~346:}}\\
       Abundances of C, N, O, Al, Si, S, Ar and Fe are in good agreement with 
       those in $\iota$~Her. P is enriched by about 
       0.15\,dex, Mg is depleted by about 0.15\,dex. Our results agree well with 
       those derived by Ryans et al. (\cite{ry96}) except for {\silii} and {\feiii}. 
       For the latter our result has a much smaller error. 
 \item {\it{PHL~159:}}\\
       Mg, Al and S are significantly depleted and O enriched by 0.3\,dex,
       whereas the other elements are in good 
       agreement with those of the comparison star.  
\end{enumerate}

Spectral analyses of massive B-type stars in open clusters as well as in the 
field (e.g. Gies \& Lambert \cite{gila92}, Kilian \cite{kili94}, Cunha \& 
Lambert \cite{cula94}) have revealed considerable variations of metal 
abundances from star to star (even within an open cluster). Kilian 
(\cite{kili94}) carried out spectral analyses of 21 B-type stars in two open 
clusters and in the field and determined abundances of C, N, O, Ne, Mg, Al,
Si, S, and Fe. We compare our results for PHL~346, BD$-$15$^\circ$115, 
PHL~159 and $\iota$~Her to her LTE results in Tables \ref{LTEphl159} and
\ref{LTEcomp}. 
Since her 
programme stars are somewhat hotter than ours, the Ne abundance is based on 
{\neii} lines, whereas we had to use {\nei} lines. 
Correcting for the significant NLTE effect on {\nei} 
(0.56~dex, see above) the neon abundance of PHL~159, the only 
programme star for which it has been measured, is found 
to be consistent with Kilian's distribution. 
The abundances we derived for all 
metals of PHL~346, BD$-$15$^\circ$115 and $\iota$~Her lie well within  
Kilian's distribution indicating that they are bona fide main sequence
B-type stars. For PHL~159, however the {\oii} abundance is higher and the {\mgii} 
abundance lower than in Kilian's distribution, whereas the other metals are 
consistent with that distribution. Therefore PHL~159 might either be a 
massive B-type star with rather peculiar abundances of the elements O and Mg or 
an evolved, low mass B-type star that mimics a massive B-type star quite closely. 
 
\section{Masses, Distances and Evolutionary Times}
The derived atmospheric parameters were compared to 
two sets of evolutionary tracks (from the Geneva group, Schaller
et al. \cite{scsc92}, and the Padua group, Salasnich et al. \cite{sagi00}) 
to estimate stellar masses and evolutionary
times {\tev} (see Fig.~\ref{schaller}) by interpolation. The results do not 
depend on the model grid used. Derived masses differ by less than 0.1\Msolar{ }and
evolutionary lifetimes by less than 4 Myr (except PG~1610$+$239: 17 Myr).  
Errors for the evolutionary lifetimes in Table~\ref{results} include errors 
propagated from uncertainties in atmospheric parameters 
as well as from the use of the two model sets. 

The distance has been calculated from mass, 
effective temperature, gravity and the dereddened apparent 
magnitude of the stars:\\[2mm]
$d = 1.11 \sqrt{\frac{M_{\star}F_V}{g}\cdot 10^{0.4V_0}}$ 
[kpc]\\[2mm] where $\rm{M_{\star}}$ is the stellar mass in $\rm{M_{\odot}}$, 
g is the gravity in cm~$\rm{s^{-2}}$, $\rm{F_V}$ is the model atmosphere flux 
at the stellar surface
in units of $10^8{\,}{\rm{erg \ cm^{-2}{ }s^{-1}{ }\AA^{-1}}}$ and $\rm{V_0}$ is 
the 
dereddened apparent visual magnitude.   

\section{Kinematics}
\subsection{Radial Velocities and Proper Motions}
Radial velocities of the slowly rotating programme stars were derived 
from the lineshift of metal lines. For the rapidly rotating stars  
only the Balmer and He {\sc i} lines could be used. 
Radial velocities obtained this way were then corrected to heliocentric 
values and the results are listed in 
Table \ref{results}. The error of the velocities estimated 
from the scatter of the velocities derived from individual lines is about 
3 -- 11 km $\rm{s^{-1}}$.Our measurements agree to within error limits with 
previous estimates (see Table~\ref{results}).\\
Proper motions were taken from literature and are listed in Table \ref{properm}.
\begin{table}
\centering
\caption[]{Data of proper motion. The position angle is
           counted positive east of north.}
\label{properm}
\begin{tabular}{|c|ccc|} \hline
 Name & $\mu$ (mas/y) & Position angle $^{\circ}$ & Reference \\ \hline
 PG~0122$+$214 & 3.4 $\pm$ 2.3 & 234 $\pm$ 71 & 1 \\ 
 PG~1533$+$467 & 16.8 $\pm$ 3.6 & 326 $\pm$ 0 & 1 \\ 
 PG~1610$+$239 & 8.1 $\pm$ 4.0 & 150 $\pm$ 16 & 1 \\ 
 PG~2219$+$094 & 6.2 $\pm$ 3.6 & 194 $\pm$ 47 & 1 \\ 
 BD$-$15$^\circ$115 & 9.0 $\pm$ 2.7 & 92 $\pm$ 10 & 2 \\ 
 PHL~346       & 8.9 $\pm$ 3.1 & 144 $\pm$ 20 & 3 \\ \hline
\end{tabular}\\[2mm]
\begin{tabular}{ll}
\parbox[t]{150mm}{\it{References}: (1) Thejll et al. (\cite{thfl97});\\ 
(2) Perryman et al. (\cite{peli97});\\ (3) Tycho-2 catalog, 
H{\o}g et al. (\cite{hofa00})}
\end{tabular}
\end{table}
\subsection{Times-of-Flight and Ejection Velocities}

The times-of-flight, which the stars need to reach their current halo 
positions from the galactic
disk, were calculated with the program ORBIT6
developed by Odenkirchen \& Brosche (\cite{odbr92}). 
This numerical code calculates the
orbit of a test body in the Galactic potential of Allen \& Santillan 
(\cite{alsa91}). The 
complete set of cylindrical coordinates is integrated and positions and
velocities are calculated in equidistant time steps. The input for this
program version are equatorial coordinates, distance d from the sun, heliocentric
radial velocities and observed absolute proper motions. Values for proper
motions are given in Table \ref{properm}. The proper motions for PHL~159,
PG~1511$+$467, SB~357 and HS~1914$+$7139 were set to zero, because
no measurements are available.
We followed the orbits backwards in time (time steps of 
0.01 --  0.1 Myr).
The time of passage through the galactic disk (= change of
sign in z--position relative to the disk) 
defines the time-of-flight \tfl. The velocity at the time of first 
crossing of the galactic plane is regarded as the ejection velocity $\rm{v_e}$
and is also calculated by the program ORBIT6.
 
Results for all parameters of the programme stars (effective
temperature, gravity, radial velocity, ejection velocity, mass, 
distance, age and time-of-flight) are summarised in Table \ref{results}. 
For the origin of the stars (see next section) the ages (\tev) 
and the times-of-flight (\tfl) are important.
We improved \tfl\ for BD$-$15$^\circ$115 and PHL~346 for which 
proper motion measurements have become available recently. 
For BD$-$15$^\circ$115 we derive a somewhat lower 
\tfl\ than Conlon et al. (\cite{codu92}) and find \tfl\ to be consistent with 
\tev\ to within our error limits. For PHL~346 we confirm that \tfl\ is 
slightly larger 
than \tev, but given the error limits this is insignificant.\\
For PG~2219$+$094 we find \tfl\ and 
\tev\ to be lower than derived by Rolleston et al. (\cite{roha99}) and \tfl\ 
to be consistent with \tev. 
For PG~1610+239 the time-of-flight is poorly constrained and only a lower 
limit could be determined which is consistent with the estimate of the 
evolutionary time.
\begin{table*}
\centering
\caption[]{Physical and kinematic 
           parameters of the programme stars. Five stars have been analysed 
           previously. 
           The results
           of these analyses from literature are given for comparison. 
           The errors of the
           evolutionary times are formal errors introduced by
           the errors in \teff{ }and \logg.}
\label{results}
\begin{tabular}{|c|ccccccccc|} \hline
 Name & $\rm{T_{eff}}$ & $\rm{\log(\frac{g}{cm s^{-2}})}$ &
 $\rm{v_{rad}}$  & $\rm{v_e}$ &
 M & d & z & $\rm{T_{flight}}$ & $\rm{T_{evol}}$  \\  
& K &  &  km s$^{-1}$ &  km s$^{-1}$ &  $\rm{M_{\odot}}$ & kpc & kpc & Myr & Myr  \\ \hline 
 PG~0122$+$214 & 18\,300 & 3.86 &  26$\pm$5 & 290 & 6.7 & 9.6 & 6.2 & 51 $\pm$ 24   & 35 $\pm$ 6 \\
                &         &      &           &     &     &     &     &               &           \\ 
 PG~1511$+$367 & 16\,100 & 4.15 & 102$\pm$11& 300:& 4.8 & 3.8 & 3.2 & 24:           & 34 $\pm$ 7  \\
                &         &      &           &     &     &     &     &               &            \\
 PG~1533$+$467 & 18\,100 & 4.00 &  33$\pm$6 & 440 & 6.0 & 3.0 & 2.4 & 20 $\pm$ 4    & 33 $\pm$ 5  \\
                &         &      &           &     &     &     &     &               &             \\
 PG~1610$+$239 & 15\,500 & 3.72 & 91$\pm$10 & 130 & 5.8 & 8.4 & 5.9 & $>$ 62 & 54 $\pm$ 10  \\
                &         &      &           &     &     &     &     &               &             \\
 PHL~159       & 18\,500 & 3.59 &  88$\pm$3 & 320:& 8.0 & 5.3 & 3.2 & 31:           & 28 $\pm$ 2  \\
                &         &      &           &     &     &     &     &               &             \\
 PG~2219$+$094  & 19\,500 & 3.58 & --24$\pm$9& 220 & 8.7 & 9.8 & 6.1 & 43 $\pm$ 22   & 27 $\pm$ 2  \\
      (1)       & 17\,900 & 3.60 &  --7      &  -- & 7.5 & --  & --  & 67            & 41          \\
                &         &      &            &    &     &     &     &               &             \\
  BD$-$15$^\circ$115 & 20\,100 & 3.81 &  93$\pm$4 &  410 & 8.0  & 4.9 & 4.8 & 30 $\pm$ 5 & 26 $\pm$ 4  \\ 
      (2)            & 19\,500 & 3.50 &  94        &  --  & 10.0 & --  & --  & 47         & 20          \\
                     &         &      &            &      &      &     &     &            &            \\
 HS~1914$+$7139     & 17\,600 & 3.90 &  --        & 330: & 6.2  & 14.9& 6.0 & 91:$^\star$  & 39 $\pm$ 6  \\ 
      (3)            & 18\,000 & 3.75 &  --39      &  --  & 6.5 -- 10.0 & 16 -- 18.4  & -- & -- & --  \\ 
                     &         &      &            &      &      &     &              &    &   \\
 PHL~346            & 20\,700 & 3.58 &  63$\pm$4  & 350  & 9.9  & 8.7 & 7.4 & 27 $\pm$ 7  & 19 $\pm$ 2 \\ 
      (4)            & 22\,600 & 3.60 &  66$\pm$10 & --   & 13.0 & --  & 8.7 & --          & 11 \\ 
                     &         &      &            &      &      &     &     &             &       \\
 SB~357             & 19\,700 & 3.90 &  58$\pm$10 & 230:  & 7.4   & 7.9   & 7.8   & 61:  & 26 $\pm$ 4\\ 
     (2)             & 19\,000 & 3.70 &  54        &  --  & 8     & --    & 9.0   & 64           & 25 \\ \hline
\end{tabular}\\[2mm]
\begin{tabular}{ll}
\parbox[t]{160mm}{\it{References}: 
(1) Rolleston et al. (\cite{roha99}); 
(2) Conlon et al. (\cite{codu92}); 
(3) Heber et al. (\cite{hemo95}); 
(4) Keenan et al. (\cite{kele86})
$^\star$ based on the radial velocity derived by Heber et al. (\cite{hemo95})\\
: uncertain due to the lack of proper motion data}
\end{tabular}\\[2mm]
\end{table*}
\section{Discussion}

We have carried out quantitative spectral analyses of ten 
apparently normal B-type stars. Their positions in the 
(\teff, \logg) diagram are consistent with models for main sequence stars. 
SB~357 shows emission in $\rm{H_{\beta}}$ and $\rm{H_{\gamma}}$, which
confirms its classification as a Be star
(Heber \& Langhans~\cite{hela86}, Kilkenny~\cite{kilk89}).
Seven of the stars have rotational velocities $>$ 70~km $\rm{s^{-1}}$
making detailed abundance analyses impossible. Mostly normal abundances with 
respect to $\iota$~Her were determined for 
BD$-$15$^\circ$115, PHL~159 and PHL~346.  
The Mg and O abundances of PHL~159 are significantly different from the 
comparison star and other normal B-type stars (Kilian, \cite{kili94}). 

Calculated orbits based on measurements of radial 
velocity and proper motion allowed to determine times-of-flight from the 
galactic plane to their present position. 

\subsection{Runaway stars}

Times-of-flight for PG~1511$+$367 and 
PG~1533$+$467 are smaller than the evolutionary times, indicating that 
these stars have been formed in the galactic 
plane and were then ejected (runaway stars). The times-of-flight are 
similar to the evolutionary times for PG~0122$+$214, 
PG~2219+094, PHL~159, BD$-$15$^\circ$115, and PG~1610+239,
which implies that the stars 
could also have formed 
in the galactic disk and were then ejected very soon after 
their birth. Ejection velocities for all programme stars range from 
130\,km s$^{-1}$ to 440\,km s$^{-1}$. 

Three mechanisms for the production of runaway stars have
been proposed in the literature:
\begin{itemize}
\item[i)] In the {\it binary supernova scenario} (Zwicky, \cite{zwi57}; Blaauw, 
\cite{bla61})
the runaway star receives its velocity after a supernova explosion in a 
massive close binary. After the explosion the binary occasionally dissociates 
and the secondary is traveling with a velocity comparable to its 
pre--explosion orbital velocity. Calculations by Iben \& Tutukov, 
(\cite{ibtu96}) indicate that the runaway star can gain a velocity of 100 to 
200\,km~s$^{-1}$.

\item[ii)] In the {\it dynamical ejection scenario} (Poveda et al., \cite{poru67};
Gies \& Bolton, \cite{gibo86}) the runaway star gains its velocity through a 
dynamical interaction with one or more stars. The most efficient encounter 
is that between two close binaries in a stellar association or open cluster, 
which in most cases results in the ejection of two runaway stars and one 
eccentric binary (Hoffer \cite{hoff83}; Mikkola \cite{mikk83}). 
Calculations by Leonard (\cite{leon91}) show that even velocities in excess 
of 1000\,km $\rm{s^{-1}}$ can be gained in rare cases.  

\item[iii)] It has been conjectured that star formation can be triggered by 
the interaction of the gas of the galactic disk with an infalling satellite 
dwarf galaxy. The momentum transferred to the gas results in a significant 
velocity component of the newly born stars perpendicular to the galactic plane.
This scenario has been supported by the discovery of the encounter with the 
Sagittarius dwarf galaxy (Ibata et al. \cite{ibat94}).  
\end{itemize} 

The availability of precise milliarcsecond astrometry for nearby stars 
through the Hipparcos satellite and pulsar astrometry and timing 
measurements have recently demonstrated that both the binary supernova 
scenario and the dynamical ejection scenario are viable. By calculating orbits 
for runaway stars, pulsars and open clusters it recently became possible to 
associate runaway stars and pulsars with their nascent clusters (Hoogerwerf et 
al., \cite{hobr00}, \cite{hobr01}). 

Our programme stars are too far away and their space motions are therefore 
not  known accurately enough to allow to identify
their relation to a young cluster or association in the galactic plane.
However, the ejection velocities determined for our programme stars may be
important to identify the mechanism which led to their ejection
from the galactic plane, once reliable theoretical predictions become 
available for the different scenarios discussed above.\\
Six stars have escape velocities exceeding 300\,km s$^{-1}$ which seems too  
large to be achievable by the binary supernova scenario.

\subsection{Stars born in the galactic halo?}

Four stars in our sample have been proposed in the literature as candidates 
for B-type stars formed in the galactic halo because their times-of-flight were
found to be considerably larger than the evolutionary time scales 
(Conlon et al., \cite{codu92}, Keenan et al., \cite {kele86}, Heber et al., 
\cite{hemo95}). As discussed above our new analysis of BD$-$15$^\circ$115
demonstrates that its time-of flight is consistent with the evolutionary 
time. Hence it could be runaway star, too.

For SB~357 and HS~1914$+$7139 the times of 
flight are 
more than twice as large as the evolutionary times, which would make 
formation in the disk unlikely.  
However, their times-of-flight are uncertain due to the lack of proper motion 
measurements. 
Such data 
are urgently needed before any firm conclusions can be drawn. 
Therefore we are reluctant to regard these stars as 
born in the halo. 

PHL~346 has been proposed as a candidate massive 
B-type star born in the halo (Ryans et al. \cite{ry96}, 
Hambly et al. \cite{hamb96}). Based on 
the new Tycho proper motion measurement, our 
analysis indicates that \tfl\ is marginally larger than \tev\
and PHL~346 can be a runaway star, too.

Hence no conclusive candidate for a young massive B-type star formed in the halo
remains in our sample. Proper motions for the four stars lacking any 
measurement should urgently be determined.  

 
\acknowledgement 
{M. R. gratefully acknowledge financial support by the DFG (grant 
He1356/27-1). We thank Michael Odenkirchen who kindly provided us with 
his code ORBIT6 for the calculation of the kinematic orbits, Heinz Edelmann who
carried out the DSAZ FOCES and ESO FEROS observations and Neil Reid, Ralf 
Napiwotzki and Klaus Werner who obtained the Keck HIRES spectra for us.}


\end{document}